\documentclass[conference,10pt]{IEEEtran}
\IEEEoverridecommandlockouts

\usepackage{epsfig,rotating,setspace,latexsym,amsmath,epsf,amssymb,amsfonts,bm,theorem,cite,algorithm,graphicx,epsf,authblk,epstopdf,color,algpseudocode,bbm}

\newtheorem{theorem}{Theorem}
\newtheorem{lemma}{Lemma}

\allowdisplaybreaks

\begin{document}

\title{Using Erasure Feedback for Online Timely Updating with an Energy Harvesting Sensor\thanks{This research was supported in part by the National Science Foundation under Grants CCF-0939370, CCF-1513915, ECCS-1650299, CNS-1526608 and ECCS-1807348.}}

\author[1]{Ahmed Arafa}
\author[2]{Jing Yang}
\author[3]{Sennur Ulukus}
\author[1]{H. Vincent Poor}
\affil[1]{\normalsize Electrical Engineering Department, Princeton University}
\affil[2]{\normalsize School of Electrical Engineering and Computer Science, The Pennsylvania State University}
\affil[3]{\normalsize Department of Electrical and Computer Engineering, University of Maryland}

\maketitle

%================================
\begin{abstract}
A real-time status updating system is considered, in which an energy harvesting sensor is acquiring measurements regarding some physical phenomenon and sending them to a destination through an {\it erasure} channel. The setting is online, in which energy arrives in units according to a Poisson process with unit rate, with arrival times being revealed causally over time. Energy is saved in a unit-sized battery. The sensor is notified by the destination of whether updates were erased via {\it feedback}. Updates need to reach the destination successfully in a {\it timely} fashion, namely, such that the long term average {\it age of information}, defined as the time elapsed since the latest successful update has reached the destination, is minimized. First, it is shown that the optimal status update policy has a {\it renewal} structure: successful update times should constitute a renewal process. Then, {\it threshold-greedy} policies are investigated: a new update is transmitted, following a successful one, only if the age of information grows above a certain threshold; and if it is erased, then all subsequent update attempts are greedily scheduled whenever energy is available. The optimal threshold-greedy policy is then analytically derived. 
\end{abstract}

%================================
\section{Introduction}

Consider a real-time status updating system, in which a destination needs to stay informed about the status of some time varying physical phenomenon through receiving time-stamped measurement updates transmitted by a sensor node. The {\it freshness} of data at the destination is captured by the {\it age of information} (AoI) metric, defined as the time elapsed since the latest update has reached the destination, and the goal is to design age-minimal status update policies that keep the information at the destination as fresh and timely as possible. However, there are three main hurdles on the way of achieving such goal: $1)$ the sensor relies on energy harvested from nature and cannot send updates all the time, $2)$ the setting is {\it online} in the sense that future energy arrivals are not known a priori, and $3)$ updates are sent through a noisy communication channel and are prone to {\it erasures}. In this work, we characterize how to optimally overcome these hurdles for the case in which the sensor is equipped with a unit-sized battery, and an erasure status {\it feedback} link exists through which the destination informs the sensor of whether its transmitted updates were successful.

There has been a plethora of works on AoI minimization in recent literature, covering topics in queuing, scheduling and coding design, e.g., \cite{yates_age_1, ephremides_age_random, chen-age-error, ephremides_age_non_linear, shroff_age_multi_hop, modiano-age-bc, sun-age-mdp, yates-age-erase-code, yates-age-cache, najm-age-multistream, najm-content-age, himanshu-age-source-coding, simeone-age-finite-code, zhong-age-source-coding}. Of particular interest, are those pertaining to energy harvesting communications \cite{yates_age_eh, elif_age_eh, arafa-age-2hop, arafa-age-var-serv, elif-age-Emax, jing-age-online, shahab-age-online-rndm, baknina-updt-info, arafa-age-online-finite, elif-age-online-threshold, jing-age-erasures-infinite-jour, arafa-age-erasure-no-fb}, in which age-minimal energy management schemes are designed. The most closely related works to this one are \cite{jing-age-erasures-infinite-jour, arafa-age-erasure-no-fb}, in which updates are transmitted through an erasure channel, with \cite{jing-age-erasures-infinite-jour} focusing on infinite battery sensors with and without erasure status feedback, and \cite{arafa-age-erasure-no-fb} focusing on unit-sized battery sensors without feedback. We note that \cite{elif_age_eh} also considers the problem without feedback, cast as an MDP, yet in a discrete time setting.

\begin{figure}[t]
\center
\includegraphics[scale=.8]{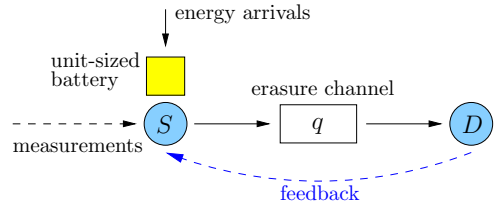}
\caption{Status updates are sent through a channel with erasure probability $q$. Feedback is received following each transmission attempt.}
%\caption{An energy harvesting sensor with a unit-sized battery sending measurement status updates to a destination through an erasure channel with erasure probability $q$. The destination sends erasure status feedback to the sensor following each transmission.}
\label{fig_sys_mod_fb}
\vspace{-.25in}
\end{figure}

In this paper, we complement the results in \cite{arafa-age-erasure-no-fb} and study the case in which the sensor gets informed by the destination of whether its transmissions were successful through a {\it feedback} link, see Fig.~\ref{fig_sys_mod_fb}. We first show that the optimal status update policy has a {\it renewal} structure, in which {\it successful} update times constitute a renewal process. Then, we focus on a class of renewal policies that we coin {\it threshold-greedy} policies. A threshold policy is one in which an update is transmitted only if the AoI grows above a certain threshold, while a greedy policy is one in which an update is transmitted once energy is available. A threshold-greedy policy combines both structures by scheduling the first update attempt, following a successful one, according to a threshold policy, and then scheduling subsequent attempts, in case the first one fails, according to a greedy policy, until the transmission is successful. We note that the existence of the feedback link is the main reason behind why the sensor can switch its policy structure after failure, unlike the indifferent threshold policy in the case without feedback in \cite{arafa-age-erasure-no-fb}. We then characterize the optimal threshold-greedy policy analytically, and show how it decreases the long term average AoI achieved in this feedback-based system, compared to that in the case without feedback in \cite{arafa-age-erasure-no-fb}.

%================================
\section{System Model and Problem Formulation} \label{sec_mod_fb}

We consider an energy harvesting sensor that is monitoring some physical phenomenon and sending measurement status updates regarding it to a destination through a noisy communication channel. Energy expenditure is normalized: one status update transmission consumes one unit of energy. The sensor is equipped with a unit-sized battery to save its incoming energy, which arrives in units according to a Poisson process with unit rate. The setting is online: future energy arrivals are known causally as they occur over time.

The effect of noise on the status updates is considered via modeling the communication channel as an erasure channel: an erasure event occurs independently for each update transmission with some probability $q\in(0,1)$, whose value is known by the sensor. Whenever update transmissions are successful, they reach the destination instantaneously within a negligible service time as in, e.g., \cite{jing-age-online, elif-age-online-threshold, arafa-age-online-finite}. A feedback link exists between the destination and the sensor, through which the sensor is informed of the erasure status following each update transmission. Such erasure feedback is sent instantaneously and error-free, and is what differentiates this work from our previous one that considered the same setting with no feedback \cite{arafa-age-erasure-no-fb}. The model considered in this paper is shown in Fig.~\ref{fig_sys_mod_fb}.

The main goal is to optimally manage the incoming energy and schedule status update transmissions such that the destination gets updated in a timely manner, namely such that long term average AoI is minimized. The AoI is mathematically defined as follows:
\begin{align}
a(t)=t-u(t),
\end{align}
where $u(t)$ is the time stamp of the most recently received update at the destination prior to time $t$.

Let $x_j$ denote the time of the $j$th transmission attempt, and let $\mathcal{E}(t)$ denote the energy available in the battery at time $t$. Therefore, {\it energy causality} dictates that
\begin{align} \label{eq_en_caus_fb}
\mathcal{E}\left(x_j^-\right)\geq1,\quad\forall j.
\end{align}
The transmission attempt times $x_j$'s must also satisfy the following battery evolution constraints:
\begin{align} \label{eq_battery_evlv_fb}
\mathcal{E}\left(x_j^-\right)=\min\left\{\mathcal{E}\left(x_{j-1}^-\right)-1+A_j,1\right\},\quad\forall j,
\end{align}
where $A_j$ is the amount of energy harvested in $[x_{j-1},x_j)$, which is, according to our (normalized) Poisson process energy arrival model, a Poisson random variable with parameter $x_j-x_{j-1}$. We assume that initially the battery is empty: $\mathcal{E}(0)=0$, and the system is fresh: $a(0)=0$.

\begin{figure}[t]
\center
\includegraphics[scale=.9]{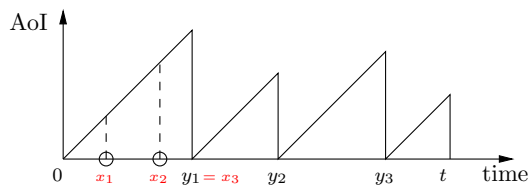}
\caption{Age evolution versus time with $n(t)=3$ successful updates. Circles denote failed attempts. In this example, the first update is successfully received after three update attempts.}
\label{fig_age_xmpl_erasure_fb}
\vspace{-.25in}
\end{figure}

Since updates are prone to erasures, we define $y_j$ as the $j$th {\it successful} update transmission time. Clearly, $\{y_j\}\subseteq\{x_j\}$. Now let $r(t)$ denote the area under the AoI evolution curve up to time $t$. This is given by
\begin{align}
r(t)=\sum_{j=1}^{n(t)}\frac{1}{2}\left(y_j-y_{j-1}\right)^2+\frac{1}{2}\left(t-y_{n(t)}\right)^2,
\end{align}
where $n(t)\triangleq\max\left\{j:~y_j<t\right\}$ denotes the number of successfully received updates by time $t$. An example of how the AoI evolves is shown in Fig.~\ref{fig_age_xmpl_erasure_fb}. Given that the sensor receives erasure feedback information, and knows the value of $q$, the problem is formulated as
\begin{align} \label{opt_main_fb}
\min_{\{x_j\}}\quad&\limsup_{T\rightarrow\infty}\frac{1}{T}\mathbb{E}\left[r(T)\right] \nonumber \\
\mbox{s.t.}\quad&(\ref{eq_en_caus_fb})-(\ref{eq_battery_evlv_fb}),
%\min_{\{x_j\}}\quad\limsup_{T\rightarrow\infty}\frac{1}{T}\mathbb{E}\left[r(T)\right] \qquad \mbox{s.t.}\quad(\ref{eq_en_caus_fb})-(\ref{eq_battery_evlv_fb}),
\end{align}
where $\mathbb{E}\left[\cdot\right]$ denotes expectation.

%================================
\section{Optimality of Renewal Policies}

The optimal $x_i$ may depend on all events (energy arrivals, erasures and update attempts) prior to $x_i$. This renders problem (\ref{opt_main_fb}) intractable. We show, however, that this is not the case, and that one can actually simplify the problem without losing optimality if we consider a fairly general class of status update policies: {\it uniformly bounded} policies, defined next. Let us denote by an {\it epoch} the time in between two consecutive successful update transmissions. A uniformly bounded policy is one in which epochs are upper bounded by a function of a finite second moment, see \cite[Definition 3]{jing-age-online}. We now state the main result of this section.

\begin{theorem} \label{thm_rnwl_fb}
The optimal uniformly bounded policy that solves problem (\ref{opt_main_fb}) is a renewal policy, in which epochs are i.i.d. and their start times $\{y_j\}$ constitute a renewal process.
\end{theorem}

We omit the proof of the theorem due to space limits. The proof, however, goes along the same lines as in \cite[Theorem~1]{arafa-age-erasure-no-fb}, where we show the optimality of renewal policies for the same system but with no feedback. The way we show it there is by considering a genie-aided setup in which a genie informs the sensor of when its updates were successful, and then argue that in the optimal policy such genie's information can be discarded. One can slightly manipulate such arguments to prove Theorem~\ref{thm_rnwl_fb} above by treating the genie-aided system as exactly the feedback system considered in this paper.

Theorem~\ref{thm_rnwl_fb} greatly simplifies problem (\ref{opt_main_fb}). It is now optimal to let the sensor {\it ignore} all the history of events once a new epoch starts (which it knows via erasure feedback), and simply repeat the same status update policy on each epoch independently. We discuss that in detail in the next section.

%================================
\section{Threshold-Greedy Policies}

\begin{figure}
\center
\includegraphics[scale=.8]{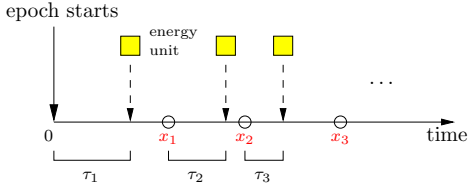}
\caption{Illustration of the notations used to describe energy arrivals and update attempt times within the epoch.}
\label{fig_tau_epoch_fb}
\vspace{-.25in}
\end{figure}

Now that the optimality of renewal-type policies is established by Theorem~\ref{thm_rnwl_fb}, we proceed with characterizing the optimal renewal policy in this section. Since epoch lengths are i.i.d., by the strong law of large numbers for renewal processes (the renewal-reward theorem) \cite{ross_stochastic} we have
\begin{align} \label{eq_obj_rnwl_fb}
\limsup_{T\rightarrow\infty}\frac{1}{T}\mathbb{E}\left[r(T)\right]=\frac{\mathbb{E}\left[R\left({\bm x}\right)\right]}{\mathbb{E}\left[L\left({\bm x}\right)\right]},
\end{align}
where $R$ denotes the area under the AoI curve (the reward) in the epoch, $L$ denotes its length, ${\bm x}=\{x_1,x_2,\dots\}$ is the update policy within the epoch where $x_i$ now denotes the time elapsed from the {\it beginning of the epoch} until the $i$th update attempt\footnote{We slightly deviate from the original definition of $x_i$ in Section~\ref{sec_mod_fb}, and assume without loss of generality that the epoch starts at time $0$.}, and the expectation is taken with respect to the energy arrivals' distribution within the epoch. Let $\tau_1$ denote the time until the first energy arrival in the epoch, and $\tau_i$, $i\geq2$, denote the time until energy arrives {\it after} the $i$th update attempt, i.e., after time $x_i$, see Fig.~\ref{fig_tau_epoch_fb}. We now have the following lemma:
\begin{lemma} \label{thm_policy_depends_aoi}
In the optimal policy, $x_i$ only depends on the AoI at $\tau_i+x_{i-1}$, i.e., $x_i\equiv x_i\left(a\left(\tau_i+x_{i-1}\right)\right)$, with $x_0\triangleq0$.
\end{lemma}
The proof of Lemma~\ref{thm_policy_depends_aoi} mainly depends on the memoryless property of the exponential distribution, along the same lines of the proof of \cite[Lemma~3]{arafa-age-online-finite}, and is omitted due to space limits. By Lemma~\ref{thm_policy_depends_aoi}, we have $x_1\equiv x_1\left(\tau_1\right)$, $x_2\equiv x_2\left(\tau_2+x_1(\tau_1)\right)$, $x_3\equiv x_3\left(\tau_3+x_2\left(\tau_2+x_1(\tau_1)\right)\right)$, and so on.

By (\ref{eq_obj_rnwl_fb}) and Lemma~\ref{thm_policy_depends_aoi}, problem (\ref{opt_main_fb}) reduces to an optimization problem over a single epoch as follows:
\begin{align} \label{opt_epoch_fb}
\min_{{\bm x}}\quad&\frac{\mathbb{E}\left[R\left({\bm x}\right)\right]}{\mathbb{E}\left[L\left({\bm x}\right)\right]} \nonumber \\
\mbox{s.t.}\quad&x_1\left(\tau_1\right)\geq\tau_1 \nonumber \\
&x_2\left(\tau_2+x_1\left(\tau_1\right)\right)\geq\tau_2+x_1\left(\tau_1\right) \nonumber \\
&x_3\left(\tau_3+x_2\left(\tau_2+x_1\left(\tau_1\right)\right)\right)\geq\tau_3+x_2\left(\tau_2+x_1\left(\tau_1\right)\right) \nonumber \\
&\dots,
\end{align}
where the inequalities represent energy causality constraints. Using iterated expectations on the (independent) erasure events, $\mathbb{E}\left[R\left({\bm x}\right)\right]$ is given by
\begin{align}
\mathbb{E}\left[R\left({\bm x}\right)\right]=&(1-q)\frac{1}{2}\mathbb{E}\left[x_1^2\left(\tau_1\right)\right] \nonumber \\
&+q(1-q)\frac{1}{2}\mathbb{E}\left[x_2^2\left(\tau_2+x_1\left(\tau_1\right)\right)\right] \nonumber \\
&+q^2(1-q)\frac{1}{2}\mathbb{E}\left[x_3^2\left(\tau_3+x_2\left(\tau_2+x_1\left(\tau_1\right)\right)\right)\right] \nonumber \\
&+\dots,
\end{align}
with $\mathbb{E}\left[L\left({\bm x}\right)\right]$ given similarly as above after excluding the $\frac{1}{2}$ terms and the squaring of the $x_i$'s.

To get a handle on problem (\ref{opt_epoch_fb}), we follow Dinkelbach's approach \cite{dinkelbach-fractional-prog} to solve this fractional program, and introduce the following parameterized auxiliary problem:
\begin{align} \label{opt_aux_fb}
p\left(\lambda\right)\triangleq\min_{{\bm x}}\quad&\mathbb{E}\left[R\left({\bm x}\right)\right]-\lambda\mathbb{E}\left[L\left({\bm x}\right)\right] \nonumber \\
\mbox{s.t.}\quad&\text{problem (\ref{opt_epoch_fb})'s constraints},
\end{align}
with $\lambda\geq0$. One can show that: $p\left(\lambda\right)$ is decreasing, and that the optimal solution of problem (\ref{opt_epoch_fb}) is given by (the unique) $\lambda^*$ that solves $p\left(\lambda^*\right)=0$ \cite{dinkelbach-fractional-prog}.

We now focus on characterizing $p(\lambda)$. Toward that, we use two terminologies in order to refer to the structure of $x_i$, for any $i$. We call $x_i$ a {\it greedy policy} if the $i$th update attempt in the epoch takes place immediately after $\tau_i$. In this case, the constraint on $x_i$ (the $i$th lower bound constraint in problem (\ref{opt_epoch_fb})) is satisfied with equality. On the other hand, we call $x_i$ a {\it $\gamma$-threshold} policy if the $i$th update attempt in the epoch only takes effect if the AoI grows above $\gamma$:
\begin{align}
x_i(t)=\begin{cases}\gamma,\quad&t<\gamma\\ t,\quad&t\geq\gamma\end{cases}.
\end{align}
We now have the following lemma (we use the notation $[\cdot]^+\triangleq\max(\cdot,0)$; the proof of the lemma is in the Appendix):
\begin{lemma} \label{thm_thrshld_grd_fb}
If $x_i$, $i\geq2$, are all greedy policies, then the optimal $x_1$ is a $\gamma$-threshold policy with $\gamma=\left[\lambda-\frac{q}{1-q}\right]^+$. Conversely, if the optimal $x_1$ is a $\gamma$-threshold policy, then the optimal $x_i$, $i\geq2$, are all greedy policies.
\end{lemma}

We coin the policies of Lemma~\ref{thm_thrshld_grd_fb} {\it threshold-greedy} policies. Employing such policies is quite intuitive in systems with feedback. Firstly, after an update is successfully transmitted, the AoI drops down to $0$. One should therefore wait for some time at least (the threshold $\gamma$ in this case) before attempting a new transmission. Such approach has been shown to be optimal in, e.g., \cite{jing-age-online, elif-age-online-threshold, arafa-age-online-finite}, in addition to the system without feedback in \cite{arafa-age-erasure-no-fb}. Secondly, if this new transmission attempt fails, then the AoI continues to increase until another energy unit arrives. It is therefore intuitive to update right away, i.e., greedily, after such energy unit arrives since the AoI is already high enough (higher than the threshold $\gamma$), and repeat that until the update is eventually successful.

Next, we focus on characterizing the optimal threshold-greedy policy by evaluating $p(\lambda)$. We basically substitute $x_1$ into equations (\ref{eq_L_final_fb}) and (\ref{eq_R_final_fb}) (see the Appendix) for two cases. First, for $\lambda<\frac{q}{1-q}$, $x_1$ is greedy, i.e., $\mathbb{E}\left[x_1(\tau_1)\right]=1$ and $\mathbb{E}\left[x_1^2(\tau_1)\right]=2$. Therefore, $p(\lambda)=1-\lambda\frac{1}{1-q}+\frac{2q-q^2}{(1-q)^2}$. Second, for $\lambda\geq\frac{q}{1-q}$, $x_1$ is a $\left(\lambda-\frac{q}{1-q}\right)$-threshold policy, and by direct computation $\mathbb{E}\left[x_1(\tau_1)\right]=\frac{1}{2}\left(\lambda-\frac{q}{1-q}\right)$ and $\mathbb{E}\left[x_1^2(\tau_1)\right]=2e^{-\left(\lambda-\frac{q}{1-q}\right)}$. Therefore, $p(\lambda)=e^{-\left(\lambda-\frac{q}{1-q}\right)}-\frac{1}{2}\lambda^2+\frac{2q-q^2}{2(1-q)^2}$. In summary, we have
\begin{align}
p(\lambda)=\begin{cases}1-\lambda\frac{1}{1-q}+\frac{2q-q^2}{(1-q)^2},\quad&\lambda<\frac{q}{1-q}\\
e^{-\left(\lambda-\frac{q}{1-q}\right)}-\frac{1}{2}\lambda^2+\frac{2q-q^2}{2(1-q)^2},\quad&\lambda\geq\frac{q}{1-q}\end{cases}.
\end{align}

We now find $\lambda^*$ that solves $p(\lambda^*)=0$. It can be directly checked that for $\lambda<\frac{q}{1-q}$, $p(\lambda)=1-\lambda\frac{1}{1-q}+\frac{2q-q^2}{(1-q)^2}>0$. Thus, focusing on the case $\lambda\geq\frac{q}{1-q}$, $\lambda^*$ is found by solving
\begin{align} \label{eq_lmda_fb}
e^{-\left(\lambda^*-\frac{q}{1-q}\right)}+\frac{2q-q^2}{2(1-q)^2}=\frac{1}{2}\left(\lambda^*\right)^2,
\end{align}
which admits a unique solution that is strictly larger than $\frac{q}{1-q}$. This can be readily verified by observing that, $1)$ the right hand side of (\ref{eq_lmda_fb}) is smaller than the left hand side for $\lambda^*=q/(1-q)$; and $2)$ the right hand side of (\ref{eq_lmda_fb}) is increasing in $\lambda^*$ while the left hand side is decreasing.

To summarize, given the erasure probability $q$, the optimal first status update policy (following a successful transmission) is a $\left(\lambda^*-\frac{q}{1-q}\right)$-threshold policy, and then all update attempts after the first one (following unsuccessful transmissions) are greedy. $\lambda^*$ is the unique solution of (\ref{eq_lmda_fb}), which also represents the long term average AoI (the value of (\ref{eq_obj_rnwl_fb})).

\begin{figure}[t]
\center
\includegraphics[scale=.4]{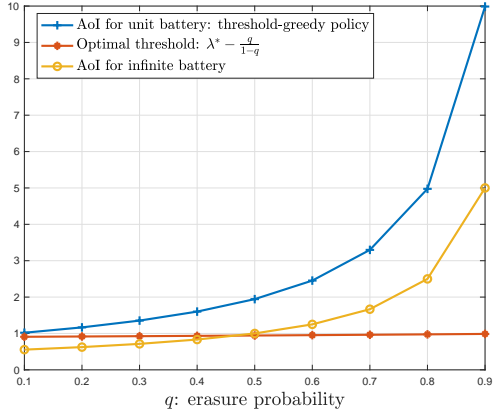}
\caption{Long term average AoI and optimal threshold vs. erasure probability for the feedback model of this paper, and that of infinite battery model \cite{jing-age-erasures-infinite-jour}.}
\label{fig_aoi_q}
\vspace{-.15in}
\end{figure}

\begin{figure}[t]
\center
\includegraphics[scale=.4]{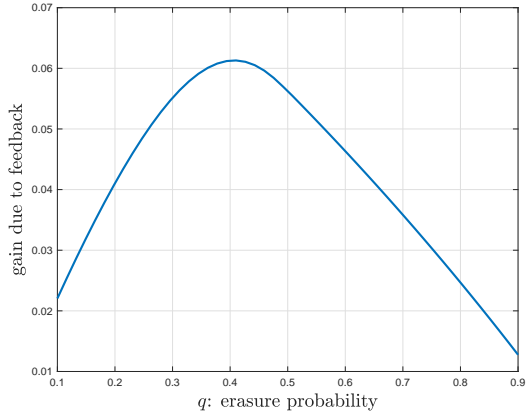}
\caption{Difference between the long term average AoI in the case without feedback \cite{arafa-age-erasure-no-fb} and that with feedback in this paper.}
\label{fig_feedback_gain}
\vspace{-.2in}
\end{figure}

%In Fig.~\ref{fig_aoi_q}, we plot the long term average AoI $\lambda^*$ versus the erasure probability $q$, under the threshold-greedy policy. We also plot the optimal threshold $\lambda^*-\frac{q}{1-q}$. We compare the results with the optimal threshold policy derived for the system with no feedback in \cite{arafa-age-erasure-no-fb}, as well as the optimal results for the infinite battery case with and without feedback derived in \cite{jing-age-error-infinite-w-fb} and \cite{jing-age-error-infinite-no-fb}, respectively. We use the variable $B$ to denote battery size.

In Fig.~\ref{fig_aoi_q}, we plot the long term average AoI $\lambda^*$ versus the erasure probability $q$. We also plot the optimal threshold $\lambda^*-\frac{q}{1-q}$, and compare the results with that of the infinite battery case, derived in \cite{jing-age-erasures-infinite-jour} to be $\frac{1}{2(1-q)}$. We see that the AoI increases with $q$, which is quite expected. We also note that the optimal threshold is almost constant. This is attributed to the fact that as $q$ increases, both $q/(1-q)$ and $\lambda^*$ from (\ref{eq_lmda_fb}) increase by almost the same amount. In Fig.~\ref{fig_feedback_gain}, we analyze the benefits of having a feedback link by plotting the difference between the long term average AoI in the system without feedback in \cite{arafa-age-erasure-no-fb} and that of this paper versus the erasure probability $q$. We denote such difference by the {\it gain due to feedback} in the figure. We observe that the gain is highest around mid values of $q$, and decreases around its extremal values. The main reason behind this is that for relatively low values of $q$, the two systems (with and without feedback) are almost identical since erasures are not very common. While for relatively high values of $q$, feedback is not really helpful since erasures would occur more frequently anyway. It is in that mid range around $q=0.4$ that feedback makes a difference.

\section{Conclusion}

The benefits of erasure status feedback has been explored for online timely updating using an energy harvesting sensor with unit-sized battery. The optimal age-minimal status update policy has been shown to have a renewal structure in which successful update times constitute a renewal process. Then, the optimal threshold-greedy policy has been characterized, in which the first update following a successful one is only transmitted if the AoI grows above a certain threshold, and then subsequent updates, in case of failure, are sent greedily whenever energy is available.

%================================
\appendix[Proof of Lemma~\ref{thm_thrshld_grd_fb}]

First, we prove the direct part: If $x_i$, $i\geq2$, are all greedy policies, then the optimal $x_1$ is a $\gamma$-threshold policy with $\gamma=\left[\lambda-\frac{q}{1-q}\right]^+$. We start by the simplifying the expected epoch length as follows:
\begin{align}
\mathbb{E}\left[L\left({\bm x}\right)\right]=&(1-q)\mathbb{E}\left[x_1(\tau_1)\right]+q(1-q)\left(1+\mathbb{E}\left[x_1(\tau_1)\right]\right) \nonumber \\
&+q^2(1-q)\left(2+\mathbb{E}\left[x_1(\tau_1)\right]\right)+\dots \nonumber \\
&+q^{i-1}(1-q)\left(i-1+\mathbb{E}\left[x_1(\tau_1)\right]\right)+\dots \\
=&\mathbb{E}\left[x_1(\tau_1)\right]+\frac{q}{1-q}. \label{eq_L_final_fb}
\end{align}
Before simplifying the expected epoch reward, let us define $G_i\triangleq\sum_{j=2}^i\tau_j$, $i\geq2$. We now proceed as follows:
\begin{align}
&\hspace{-.15in}\mathbb{E}\left[R\left({\bm x}\right)\right] \nonumber \\
%=&(1-q)\frac{1}{2}\mathbb{E}\left[x_1^2(\tau_1)\right]+q(1-q)\frac{1}{2}\mathbb{E}\left[\left(\tau_2+x_1(\tau_1)\right)^2\right] \nonumber \\
%&+q^2(1-q)\mathbb{E}\left[\left(\tau_3+\tau_2+x_1(\tau_1)\right)^2\right]+\dots \nonumber \\
%&+q^{i-1}(1-q)\mathbb{E}\left[\left(\tau_i+\tau_{i-1}+\dots+\tau_2+x_1(\tau_1)\right)^2\right]+\dots \\
=&(1-q)\frac{1}{2}\mathbb{E}\left[x_1^2(\tau_1)\right] +\sum_{i=2}^\infty q^{i-1}(1-q)\mathbb{E}\left[\left(G_i+x_1(\tau_1)\right)^2\right] \\
=&(1-q)\frac{1}{2}\mathbb{E}\left[x_1^2(\tau_1)\right] \nonumber \\
%&+q(1-q)\left(\frac{1}{2}\mathbb{E}\left[\tau_2^2\right]+\frac{1}{2}\mathbb{E}\left[x_1^2(\tau_1)\right]+\mathbb{E}\left[\tau_2\right]\mathbb{E}\left[x_1(\tau_1)\right]\right) \nonumber \\
%&+q^2(1-q)\left(\!\frac{1}{2}\mathbb{E}\left[G_3^2\right]\!+\!\frac{1}{2}\mathbb{E}\left[x_1^2(\tau_1)\right]\!+\!\mathbb{E}\left[G_3\right]\mathbb{E}\left[x_1(\tau_1)\right]\!\right) \nonumber \\
%&+\dots \nonumber \\
%&+q^{i-1}(1-q)\mathbb{E}\left(\frac{1}{2}\mathbb{E}\left[G_i^2\right]+\frac{1}{2}\mathbb{E}\left[x_1^2(\tau_1)\right]\right. \nonumber \\
%&\hspace{2in}+\mathbb{E}\left[G_i\right]\mathbb{E}\left[x_1(\tau_1)\right]\bigg) \nonumber \\
%&+\dots \\
&+\sum_{i=2}^\infty q^{i-1}(1-q)\mathbb{E}\left(\frac{1}{2}\mathbb{E}\left[G_i^2\right]+\frac{1}{2}\mathbb{E}\left[x_1^2(\tau_1)\right]\right. \nonumber \\
&\hspace{1.75in}+\mathbb{E}\left[G_i\right]\mathbb{E}\left[x_1(\tau_1)\right]\bigg) \\
=&\frac{1}{2}\mathbb{E}\left[x_1^2(\tau_1)\right]+\frac{q}{1-q}\mathbb{E}\left[x_1(\tau_1)\right] \nonumber \\
&+\frac{1}{2}\sum_{i=1}^\infty\left(i-1+(i-1)^2\right)q^{i-1}(1-q) \label{eq_smplfy_R_fb} \\
=&\frac{1}{2}\mathbb{E}\left[x_1^2(\tau_1)\right]+\frac{q}{1-q}\mathbb{E}\left[x_1(\tau_1)\right]+\frac{q}{(1-q)^2}, \label{eq_R_final_fb}
\end{align}
where (\ref{eq_smplfy_R_fb}) follows by the fact that that $G_i$ has a gamma distribution with parameters $i-1$ and $1$, and, in particular, its second moment is given by $\mathbb{E}\left[G_i^2\right]=i-1+(i-1)^2$.

We now plug in (\ref{eq_L_final_fb}) and (\ref{eq_R_final_fb}) into the objective function of problem (\ref{opt_aux_fb}), and introduce the following Lagrangian \cite{boyd}:
\begin{align}
\mathcal{L}=&\frac{1}{2}\mathbb{E}\left[x_1^2(\tau_1)\right]+\left(\frac{q}{1-q}-\lambda\right)\mathbb{E}\left[x_1(\tau_1)\right]+\frac{q}{(1-q)^2} \nonumber \\
&-\lambda\frac{q}{1-q}-\int_0^\infty\eta_1(\tau_1)\left(x_1(\tau_1)-\tau_1\right)d\tau_1,
\end{align}
where $\eta_1$ is a Lagrange multiplier. Taking the (functional) derivative with respect to $x_1(t)$ and equating to $0$ we get
\begin{align}
x_1(t)=\left(\lambda-\frac{q}{1-q}\right)+\frac{\eta_1(t)}{e^{-t}}.
\end{align}
We now have two cases. The first is when $\lambda<\frac{q}{1-q}$, whence $\eta_1(t)$ must be strictly positive $\forall t$, which implies by complementary slackness \cite{boyd} that $x_1(t)=t,~\forall t$. In other words, $x_1$ in this case is a greedy policy, or equivalently a $0$-threshold policy. The second case is when $\lambda\geq\frac{q}{1-q}$, in which similar analysis to that in \cite[Section~3]{arafa-age-online-finite} can be carried out to show that $x_1$ is a $\left(\lambda-\frac{q}{1-q}\right)$-threshold policy. Combining both cases concludes the proof of the direct part.

We now prove the converse part: if the optimal $x_1$ is a $\gamma$-threshold policy, then the optimal $x_i$, $i\geq2$, are all greedy policies. Hence, the first update attempt occurs optimally (by hypothesis) at $x_1(\tau_1)$. Assume that it fails. Note that, by construction, $\tau_2>x_1(\tau_1)$ (see Fig.~\ref{fig_tau_epoch_fb}). Let $s_2\triangleq\tau_2+x_1(\tau_1)$, and let $x_2$ be {\it not} greedy: $x_2(s_2)=s_2^\prime$ for some $s_2^\prime>s_2$. Now consider a slightly different energy arrival pattern, in which the first energy arrival occurs at $s_2$, as opposed to $\tau_1$. Since $s_2>x_1(\tau_1)$, and $x_1$ is an optimal threshold policy, therefore it holds that $x_1(s_2)=s_2$, i.e., it is optimal to update right away at time $s_2$ in the second sample path situation.

Now observe that in both situations the AoI $a(s_2)=s_2$; and, by the memoryless property of exponential distribution, that the time until the next energy arrival after $s_2$ is $\sim\exp(1)$. In addition, the probability that an update gets erased is independent of past erasures. Given that $a(s_2)=s_2$, the upcoming energy arrival is $\sim\exp(1)$, and the probability of erasure is $q$, the optimal decision in the second situation is $x_1(s_2)=s_2$, i.e., update exactly at $s_2$. Therefore, in the first situation, in which the same statistical conditions hold at $s_2$, it {\it cannot be optimal} to wait and update at time $s_2^\prime$. Hence, $x_2$ must be greedy. Similar arguments hold to show that $x_i$, $i\geq3$, must all be greedy as well, given that the optimal $x_1$ is a threshold policy. This concludes the proof of the converse part, and that of the lemma.

%================================


\begin{thebibliography}{10}

\bibitem{yates_age_1}
S.~Kaul, R.~D. Yates, and M.~Gruteser.
\newblock Real-time status: How often should one update?
\newblock In {\em Proc. IEEE Infocom}, March 2012.

\bibitem{ephremides_age_random}
C.~Kam, S.~Kompella, and A.~Ephremides.
\newblock Age of information under random updates.
\newblock In {\em Proc. IEEE ISIT}, July 2013.

\bibitem{chen-age-error}
K.~Chen and L.~Huang.
\newblock Age-of-information in the presence of error.
\newblock In {\em Proc. IEEE ISIT}, June 2016.

\bibitem{ephremides_age_non_linear}
A.~Kosta, N.~Pappas, A.~Ephremides, and V.~Angelakis.
\newblock Age and value of information: Non-linear age case.
\newblock In {\em Proc. IEEE ISIT}, June 2017.

\bibitem{shroff_age_multi_hop}
A.~M. Bedewy, Y.~Sun, and N.~B. Shroff.
\newblock Age-optimal information updates in multihop networks.
\newblock In {\em Proc. IEEE ISIT}, June, 2017.
\newblock Longer version available: ar{X}iv:1712.10061.

\bibitem{modiano-age-bc}
Y.~Hsu, E.~Modiano, and L.~Duan.
\newblock Age of information: Design and analysis of optimal scheduling
  algorithms.
\newblock In {\em Proc. IEEE ISIT}, June 2017.

\bibitem{sun-age-mdp}
Y.~Sun, E.~Uysal-Biyikoglu, R.~D. Yates, C.~E. Koksal, and N.~B. Shroff.
\newblock Update or wait: How to keep your data fresh.
\newblock {\em IEEE Trans. Inf. Theory}, 63(11):7492--7508, November 2017.

\bibitem{yates-age-erase-code}
R.~D. Yates, E.~Najm, E.~Soljanin, and J.~Zhong.
\newblock Timely updates over an erasure channel.
\newblock In {\em Proc. IEEE ISIT}, June 2017.

\bibitem{yates-age-cache}
R.~D. Yates, P.~Ciblat, A.~Yener, and M.~A. Wigger.
\newblock Age-optimal constrained cache updating.
\newblock In {\em Proc. IEEE ISIT}, June 2017.

\bibitem{najm-age-multistream}
E.~Najm and E.~Telatar.
\newblock Status updates in a multi-stream {M}/{G}/1/1 preemptive queue.
\newblock In {\em Proc. IEEE Infocom}, April 2018.

\bibitem{najm-content-age}
E.~Najm, R.~Nasser, and E.~Telatar.
\newblock Content based status updates.
\newblock In {\em Proc. IEEE ISIT}, June 2018.

\bibitem{himanshu-age-source-coding}
P.~Mayekar, P.~Parag, and H.~Tyagi.
\newblock Optimal lossless source codes for timely updates.
\newblock In {\em Proc. IEEE ISIT}, June 2018.

\bibitem{simeone-age-finite-code}
R.~Devassy, G.~Durisi, G.~C. Ferrante, O.~Simeone, and E.~Uysal-Biyikoglu.
\newblock Delay and peak-age violation probability in short-packet
  transmissions.
\newblock In {\em Proc. IEEE ISIT}, June 2018.

\bibitem{zhong-age-source-coding}
J.~Zhong, R.~D. Yates, and E.~Soljanin.
\newblock Timely lossless source coding for randomly arriving symbols.
\newblock In {\em Proc. ITW}, November 2018.

\bibitem{yates_age_eh}
R.~D. Yates.
\newblock Lazy is timely: Status updates by an energy harvesting source.
\newblock In {\em Proc. IEEE ISIT}, June 2015.

\bibitem{elif_age_eh}
B.~T. Bacinoglu, E.~T. Ceran, and E.~Uysal-Biyikoglu.
\newblock Age of information under energy replenishment constraints.
\newblock In {\em Proc. ITA}, February 2015.

\bibitem{arafa-age-2hop}
A.~Arafa and S.~Ulukus.
\newblock Age-minimal transmission in energy harvesting two-hop networks.
\newblock In {\em Proc. IEEE Globecom}, December 2017.

\bibitem{arafa-age-var-serv}
A.~Arafa and S.~Ulukus.
\newblock Age minimization in energy harvesting communications:
  Energy-controlled delays.
\newblock In {\em Proc. Asilomar}, October 2017.

\bibitem{elif-age-Emax}
B.~T. Bacinoglu and E.~Uysal-Biyikoglu.
\newblock Scheduling status updates to minimize age of information with an
  energy harvesting sensor.
\newblock In {\em Proc. IEEE ISIT}, June 2017.

\bibitem{jing-age-online}
X.~Wu, J.~Yang, and J.~Wu.
\newblock Optimal status update for age of information minimization with an
  energy harvesting source.
\newblock {\em IEEE Trans. Green Commun. Netw.}, 2(1):193--204, March 2018.

\bibitem{shahab-age-online-rndm}
S.~Farazi, A.~G. Klein, and D.~R.~Brown III.
\newblock Average age of information for status update systems with an energy
  harvesting server.
\newblock In {\em Proc. IEEE Infocom}, April 2018.

\bibitem{baknina-updt-info}
A.~Baknina, O.~Ozel, J.~Yang, S.~Ulukus, and A.~Yener.
\newblock Sending information through status updates.
\newblock In {\em Proc. IEEE ISIT}, June 2018.

\bibitem{arafa-age-online-finite}
A.~Arafa, J.~Yang, S.~Ulukus, and H.~V. Poor.
\newblock Age-minimal transmission for energy harvesting sensors with finite
  batteries: Online policies.
\newblock Available Online: ar{X}iv:1806.07271.

\bibitem{elif-age-online-threshold}
B.~T. Bacinoglu, Y.~Sun, E.~Uysal-Biyikoglu, and V.~Mutlu.
\newblock Achieving the age-energy tradeoff with a finite-battery energy
  harvesting source.
\newblock In {\em Proc. IEEE ISIT}, June 2018.

\bibitem{jing-age-erasures-infinite-jour}
S.~Feng and J.~Yang.
\newblock Age of information minimization for an energy harvesting source with
  updating erasures: Without and with feedback.
\newblock Available Online: ar{X}iv1808.05141.

\bibitem{arafa-age-erasure-no-fb}
A.~Arafa, J.~Yang, S.~Ulukus, and H.~V. Poor.
\newblock Online timely status updates with erasures for energy harvesting
  sensors.
\newblock In {\em Proc. 56th Annu. Allerton Conf. Commun. Contr. Comput.},
  October 2018.

\bibitem{ross_stochastic}
S.~M. Ross.
\newblock {\em Stochastic Processes}.
\newblock Wiley, 1996.

\bibitem{dinkelbach-fractional-prog}
W.~Dinkelbach.
\newblock On nonlinear fractional programming.
\newblock {\em Management Science}, 13(7):492--498, 1967.

\bibitem{boyd}
S.~P. Boyd and L.~Vandenberghe.
\newblock {\em Convex Optimization}.
\newblock Cambridge University Press, 2004.

\end{thebibliography}
\end{document}